\documentclass[aps,prl,twocolumn,superscriptaddress,showpacs]{revtex4}

\usepackage[dvips]{graphicx}
\usepackage{amsmath,amssymb,latexsym,bm}

\begin{document}

%\preprint{}

\title{
Dynamical band flipping in fermionic lattice systems:
An ac-field-driven change of the interaction from repulsive to attractive
%Changing the interaction of lattice fermions dynamically\\ 
%from repulsive to attractive in ac fields
} 

\author{Naoto Tsuji}
\affiliation{Department of Physics, University of Tokyo, Hongo, Tokyo 113-0033, Japan}
\author{Takashi Oka}
\affiliation{Department of Physics, University of Tokyo, Hongo, Tokyo 113-0033, Japan}
\author{Philipp Werner}
\affiliation{Theoretical Physics, ETH Zurich, 8093 Z\"{u}rich, Switzerland}
\author{Hideo Aoki}
\affiliation{Department of Physics, University of Tokyo, Hongo, Tokyo 113-0033, Japan}

%\email[]{}
%\homepage[]{}
%\thanks{}
%\altaffiliation{}
%\collaboration{}
%\noaffiliation

\date{\today}

\begin{abstract}
%We show theoretically that a sudden application of 
%an appropriate ac field flips a band structure,
%which effectively switches the interfermion interaction from repulsive to 
%attractive 
%in correlated fermion systems on a lattice. 
%The nonadiabatically driven system is characterized by a negative temperature 
%with a population inversion.   
%We numerically demonstrate the dynamical band flipping 
%in an ac-driven Hubbard model 
%with the nonequilibrium dynamical mean-field theory solved by the 
%continuous-time quantum Monte Carlo method.  
%Based on this, we propose a possibility that an ac field can dynamically induce 
%superconductivity if it is turned on in a suitable manner.
We show theoretically that the sudden application of an appropriate ac field 
to correlated lattice fermions flips the band structure and 
effectively switches the interaction from repulsive to attractive. 
The nonadiabatically driven system is characterized by a negative temperature 
with a population inversion.   
We numerically demonstrate the converted interaction 
in an ac-driven Hubbard model 
with the nonequilibrium dynamical mean-field theory solved by the 
continuous-time quantum Monte Carlo method.  
Based on this, we propose 
%the possibility that an ac field can dynamically induce 
%superconductivity if it is turned on in a suitable manner.
an efficient ramp-up protocol for ac fields 
that can suppress heating, which leads to an effectively attractive 
Hubbard model with a temperature below the superconducting transition
temperature of the equilibrium system.
%low enough that a superconducting state is expected to emerge.
\end{abstract}

%\keywords{}

%03.75.Lm    : Tunneling, Josephson effect, Bose-Einstein condensates in periodic potentials, 
%              solitons, vortices, and topological excitations
%03.75.Ss    : Matter waves
%                --- Degenerate Fermi gases 
%03.75.-b    : Matter waves
%7.85.-d     : Ultracold gases, trapped gases
%67.85.Lm    : Degenerate Fermi gases 
%03.65.Vf    : Quantum mechanics
%                --- Phases: geometric; dynamic or topological 
%71.10.Fd    : Theories and models of many-electron systems
%                --- Lattice fermion models (Hubbard model, etc.)
%05.70.Ln    : Thermodynamics
%                --- Nonequilibrium and irreversible thermodynamics
%05.10.Ln    : Computational methods in statistical physics and nonlinear dynamics
%                --- Monte Carlo methods
%71.27.+a   : Strongly correlated electron systems; heavy fermions
%05.30.Fk   : Quantum statistical mechanics 
%               --- Fermion systems and electron gas

\maketitle

{\it Introduction.---} 
%\label{intro}
There is an increasing fascination with 
the dynamics of fermions driven by external fields as a result of 
recent developments in time-resolved experimental techniques
in, e.g., ultracold atom physics \cite{BlochDalibardZwerger2008}
and electrons in a crystal \cite{Cavalieri2007}.
There, the external fields are employed not only for probing the response against perturbation, 
but also for creating excited states to control phases of the system.
In particular, continuously driven systems isolated from the 
environment 
exhibit nonequilibrium statistical distributions that 
dramatically alter their physical properties.
One basic, long-known example is the 
negative temperature ($T$) distribution \cite{RamseyKlein1956}, where
higher energy levels are occupied (population inversion).
If such distributions are realized in 
correlated fermion systems, 
it should have a huge impact on many-body physics since,
as we shall show, 
the originally repulsive interfermion interaction effectively turns 
into an attractive one as a result of the negative $T$.

To start with, a realization of negative $T$ requires as a crucial condition 
that the energy spectrum is upper bounded \cite{RamseyKlein1956}.
%In fact Ref.~\cite{Ramsey1956}, 
A spin system, for which the concept of negative $T$ was originally introduced \cite{PurcellPound1951},
satisfies this condition.
%and a three or four level system, a model for lasers \cite{SchawlowTownes1958}, are 
Another promising candidate for negative $T$ is a many-particle system on a lattice, 
where particles form a band structure with a finite bandwidth, 
as far as the interband excitations induced by external fields 
are forbidden or negligible.
For instance, Ref.~\cite{Mosk2005} proposes that a harmonic potential
trapping bosonic atoms in an optical lattice can be inverted to realize a negative $T$.

In the present Letter, we take a totally different approach, namely, 
a sudden application (quench) of a sinusoidal force (ac field)
to a lattice fermion system.
The effect of ac fields has been discussed previously in the context of 
adiabatic processes (e.g., \cite{EckardtWeissHolthaus2005}), where the hopping
amplitude is effectively renormalized \cite{DunlapKenkreHolthaus}. 
Here we show that the ``nonadiabatic'' switch-on of the ac field can
dynamically invert the band structure (dynamical band flipping),
%due to the hopping renormalization, 
%We shall show that a sudden application (quench) 
%of a sinusoidal force (ac field) may
generating a negative-$T$ distribution. 
%in a lattice fermion system 
%where the band structure is {\it dynamically inverted}.  
%We suggest that the ac-quench should be related to the interaction-quench \cite{EcksteinKollarWerner2009} (ac/$U$-quench correspondence),
%which implies that the interaction is effectively converted between 
%repulsive and attractive.
%This is caused by an effective renormalization 
%of a hopping amplitude in the presence of ac fields, which is known to hold 
%in adiabatic processes (e.g., \cite{EckardtWeissHolthaus2005}), but here shown to be also the case in {\it nonadiabatic} situations. 
Based on this, we suggest that the ac quench is related to an interaction quench (``ac/$U$-quench correspondence''), 
where the interaction can be effectively converted between repulsive and attractive.
This is numerically demonstrated 
%by an enhanced double occupancy and the inverted distribution function in
for the ac-driven Hubbard model, 
whose time evolution is obtained with the nonequilibrium dynamical mean-field theory (DMFT) 
\cite{GeorgesKotliarKrauthRozenberg1996a, FreericksTurkowskiZlatic2006}
solved with the continuous-time quantum Monte Carlo (QMC) method \cite{WernerOkaMillis2009}.  
%The result indicates that the 
%interaction is effectively converted from repulsive to attractive, 
%which may first seem counterintuitive but possible for a negative $T$, 
%as confirmed from the distribution function and 
%a comparison with $U$-quench data.  

One immediate question may be whether the sudden switch-on of the 
ac field causes a violent heating of the system. 
However, we shall show that the heating associated with the sign change of the interaction
can be suppressed if the ac field is turned on in a suitable way.
Therefore, if the ac-driven system thermalizes,
%In particular, the ac-driven system, if thermalizes, can finally arrive at 
it will correspond to
an effectively attractive Hubbard model whose temperature is 
low enough that the system can possibly accommodate 
superconductivity.
%a careful choice in the way in which the ac field is 
%turned on can suppress the heating.
%To be more precise, what we want is a sharp Fermi surface 
%(with the momentum distribution close to the step function), 
%and a sharper Fermi surface corresponds, for negative $T$, 
%a smaller $|T|$, which can be realised for the appropriate switch-on 
%of the ac field.   (Note that the sequence from low $T$ to high $T$ is 
%$0\to \infty \to -\infty \to -0$.)  
%By combining the repulsive/attractive conversion and the 
%moderate heating, we shall finally propose that the ac-driven 
%system can possibly accommodate 
%superconductivity arising from the effectively attractive interaction, 
%which by its nature should have a higher $T_c$ than 
%in the repulsive case.  

{\it Formulation.---} 
We consider a model Hamiltonian, 
\begin{align}
  {\cal H}(t)
    =
      -J \mathcal{H}_{K}+U \mathcal{H}_{I}
      +\cos(\Omega t) \sum_{j,\sigma} \bm{K}(t)\cdot \bm{R}_j n_{j\sigma}
  \label{hubbard}
%  \\
%  &\mathcal{H}_{K}
%    =
%    \sum_{\langle ij \rangle,\sigma} c_{i\sigma}^\dagger c_{j\sigma},
%    \sum_{\langle ij \rangle,\sigma} (c_{i\sigma}^\dagger c_{j\sigma}+{\rm h.c.}),
%    \;
%  \mathcal{H}_{I}
%    =
%    \sum_{j}\left(n_{j\uparrow}-\frac{1}{2}\right)\left(n_{j\downarrow}-\frac{1}{2}\right),
%    \sum_{j} n_{j\uparrow}n_{j\downarrow},
%  \nonumber
\end{align}
with $ \mathcal{H}_{K}=\sum_{\langle ij \rangle,\sigma} (c_{i\sigma}^\dagger c_{j\sigma}+{\rm H.c.})$ and
$\mathcal{H}_{I}=\sum_{j}\left(n_{j\uparrow}-\frac{1}{2}\right)\left(n_{j\downarrow}-\frac{1}{2}\right)$.
Here $c_{j\sigma}^\dagger$ creates a fermion at site $j$ with spin $\sigma$ for electron systems 
or pseudospin $\sigma$ for cold atoms, 
${\bm R}_j$ is a position vector, 
$n_{j\sigma}=c_{j\sigma}^\dagger c_{j\sigma}$,
$J(>0)$ the hopping energy, $U(>0)$ the on-site interaction.   
The third term in Eq.~(\ref{hubbard}) represents the ac field with amplitude 
${\bm K}$ and frequency $\Omega$.
We take, for the DMFT, 
a $d$-dimensional hypercubic lattice with ${\bm K}(t)=K(t)(1,1,\dots,1)$,
and consider a half-filled band.
%for which the chemical potential is included in (\ref{hubbard}).
We assume that the system is initially in equilibrium at temperature $T$, 
and the ac field is suddenly switched on at $t=0$,
%and is being applied to the system continuously after that, 
i.e., $K(t)=K\theta(t)$ with $\theta(t)$ the step function. 

\begin{figure}[t]
%  \begin{center}
%    \begin{tabular}{cc}
%      \hspace{-3cm}
%      \includegraphics[width=105mm]{dbleocc1.eps} &
%      \hspace{-6.5cm}
%      \includegraphics[width=105mm]{dbleocc2.eps} \\
%    \end{tabular}
    \includegraphics[width=10cm]{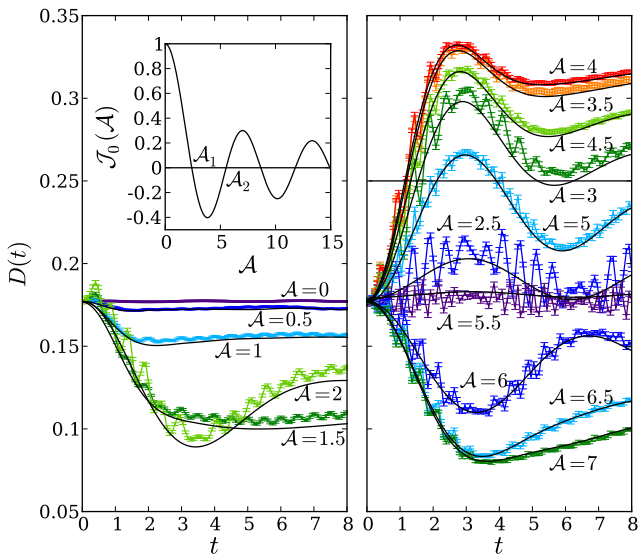}
    \caption{Time evolution of the double occupancy $D$ for various 
    values of $\mathcal{A}\equiv K/\Omega$ (symbols with error bars) for $\Omega=2\pi$, $U=1$, 
    with the horizontal line indicating the free-electron value of $D=0.25$.  
    Solid curves represent the corresponding result for the $U$ quench with $\rho_{\mathcal{A}}^+(\tilde{t})$ 
    plotted on a rescaled time axis $t=\tilde{t}/|\mathcal{J}_0(\mathcal{A})|$. 
    Inset depicts the Bessel function, $\mathcal{J}_0(\mathcal{A})$.}
    \label{dbleocc}
%  \end{center}
\end{figure}

It has been shown \cite{DunlapKenkreHolthaus}, for 
noninteracting systems, that the effect of the ac field is simply a 
renormalization of the hopping energy, 
\begin{align}
  J
    \to
      J_{\rm eff}=\mathcal{J}_0(K/\Omega)J,
  \label{bessel}
\end{align}
with $\mathcal{J}_n(z)$ the $n$th order Bessel function. 
A naive explanation of  Eq.~(\ref{bessel}) is that in the presence of the ac field
the original band dispersion $\epsilon_{\bm k}=-2J\sum_{i=1}^d \cos k_i$ is replaced with
a time-averaged 
$\overline{\epsilon_{\bm k}}=\frac{\Omega}{2\pi}\int_0^{2\pi/\Omega} dt \epsilon_{{\bm k}-{\bm A}(t)}
=\mathcal{J}_0(K/\Omega)\epsilon_{\bm k}$,
%by Floquet theory \cite{Shirley1965, Sambe1973}
%which provides a useful argument about periodically driven systems. 
%Let $h(t)$ be the noninteracting part of the Hamiltonian (the first and third terms of Eq.~(\ref{hubbard})).
where we take a gauge in which the ac field is represented by a vector potential 
${\bm A}(t)=-{\bm K}\sin(\Omega t)/\Omega$. 
The renormalization of $J$ can more rigorously 
be derived from the Floquet theory for ac fields \cite{TsujiOkaAoki2008}.
%we can derive the Floquet matrix for the Hamiltonian
%The eigenvalues of $\hat{h}$ give the single-particle spectrum of the noninteracting system.
%Since $\hat{h}$ can be exactly diagonalized with the eigenvalues
%$\hat{h}_{nn}=\mathcal{J}_0(\mathcal{A})\epsilon_{\bm k}-n\Omega$
%for any $\Omega$ and $\mathcal{A}$ \cite{TsujiOkaAoki2008}, 
%the spectrum has Floquet replicas with the effective renormalization (\ref{bessel})
%of the band width and a constant shift $n\Omega$. If $\Omega\gg J$
%Floquet sidebands ($n \neq 0$) are negligible,
%and the system is essentially described by $\hat{h}_{00}=\mathcal{J}_0(\mathcal{A})\epsilon_{\bm k}$, 
%leading to (\ref{bessel}).
%Effects of many-body interactions on the scaling (\ref{bessel}) have been studied
%for the semiconductor superlattice model \cite{MeiervonPlessenThomasKoch1995},
%the Bose-Hubbard model \cite{EckardtWeissHolthaus2005, CreffieldMonteiro2006}, 
%and the Falicov-Kimball model \cite{TsujiOkaAoki2008}.
Curiously, $J_{\rm eff}\propto\mathcal{J}_0(K/\Omega)$ 
then 
%is an oscillating function, $J_{\rm eff}$ 
changes sign as $\mathcal{A}\equiv K/\Omega$ is increased. 
This implies that $J_{\rm eff}$ vanishes at $\mathcal{A}_1=2.404\dots$ (see the inset of Fig.~\ref{dbleocc}), which 
is known as dynamical localization \cite{DunlapKenkreHolthaus}. 
%or, in the context of two-level systems, the 
%coherent destruction of tunneling \cite{GrossmannDittrichJungHanggi1991a}.
Experimentally, the scaling (\ref{bessel}) was beautifully 
observed in a Bose-Einstein condensate (BEC) in an optical lattice \cite{Lignier2007}.

What happens when $\mathcal{J}_0(\mathcal{A})<0$? 
In equilibrium, the inverted sign of $J$ does not change the physics, 
since it can be canceled with a gauge transformation 
$c_{j\sigma}^\dagger \to -c_{j\sigma}^\dagger$ for the $j\in B$ sublattice.
%the Fermi distribution just follows the change in the band dispersion.
Out of equilibrium, however, we have to consider the distribution of occupied states,
%nonequilibrium distributions (not necessarily equal to equilibrium Fermi distribution
%$f_T(\epsilon_{\bm k})=1/(e^{\epsilon_{\bm k}/T}+1)$) are considered, 
and the physics can indeed dramatically
change as we cross the zeros of $\mathcal{J}_0(\mathcal{A})$.

\begin{figure}[b]
  \begin{center}
    \includegraphics[width=6cm]{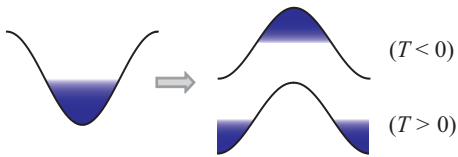}
    \caption{A schematic band flipping accompanied with 
    a population inversion ($T<0$) induced by the ac field 
    when $\mathcal{J}_0(\mathcal{A})<0$ and without a population inversion ($T>0$).}
%    \vspace{-.4cm}
    \label{inversion}
  \end{center}
\end{figure}
To characterize nonequilibrium distributions, we can define a time-resolved spectral function, 
$A(\omega,t)\equiv -\frac{1}{\pi}{\rm Im}\int dt' e^{i\omega t'}\sum_{\bm k}
G_{{\bm k}\sigma}^R(t+\frac{t'}{2},t-\frac{t'}{2})$ and an occupation 
$N(\omega,t)\equiv -\frac{i}{2\pi}\int dt' e^{i\omega t'}\sum_{\bm k} 
G_{{\bm k}\sigma}^<(t+\frac{t'}{2},t-\frac{t'}{2})$, 
in terms of the retarded $G_{{\bm k}\sigma}^R(t,t')=
-i\theta(t-t')\langle \{c_{{\bm k}\sigma}(t),c_{{\bm k}\sigma}^\dagger(t')\} \rangle$
and lesser Green function $G_{{\bm k}\sigma}^<(t,t')=
i\langle c_{{\bm k}\sigma}^\dagger(t)c_{{\bm k}\sigma}(t')\rangle$.
%Each function represents density of energy levels with energy $\omega$ %and density of occupied levels at time $t$. 
%For noninteracting fermions, the retarded Green function reads
%$G_{{\bm k}\sigma}^{R0}(t,t')=-i\theta(t-t')
%e^{-i\!\int_{t'}^{t} d\bar{t} \epsilon_{{\bm k}-{\bm A}(\bar{t})}}$,
%while the lesser one is
%$G_{{\bm k}\sigma}^{<0}(t,t')=if_T(\epsilon_{\bm k})
%e^{-i\!\int_{t'}^{t} d\bar{t} \epsilon_{{\bm k}-{\bm A}(\bar{t})}}$.
For noninteracting fermions we can evaluate \cite{TsujiOkaAoki2008}, 
for $\Omega\gg J$, the longtime behavior as
\begin{align}
  A(\omega,t)
    &\xrightarrow{t\to\infty}
      \frac{1}{|\mathcal{J}_0(\mathcal{A})|}\rho_0\!\left(\frac{\omega}{|\mathcal{J}_0(\mathcal{A})|}\right),
  \label{spectrum}
  \\
  N(\omega,t)
    &\xrightarrow{t\to\infty}
      A(\omega,\infty)f_T\!\left(\frac{\omega}{\mathcal{J}_0(\mathcal{A})}\right),
  \label{occupation}
\end{align}
where $\rho_0(\omega)=\sum_{\bm k}\delta(\omega-\epsilon_{\bm k})$ is the noninteracting density of states,
$f_T(\omega)=1/(e^{\omega/T}+1)$ the 
Fermi distribution with $k_B=1$,
and we have used the symmetry $\rho_0(-\omega)=\rho_0(\omega)$.
%The asympotic forms (\ref{spectrum}) and (\ref{occupation}) also contain
%residual oscillating terms proportional to $e^{in\Omega t}$, 
%which are out of interest here.
%but here we focus on dc components. 
%From (\ref{spectrum}) and (\ref{occupation}), 
We notice that the sign of $\mathcal{J}_0(\mathcal{A})$ is irrelevant to the spectrum, 
but it is indeed relevant to the occupation. In fact, for $\mathcal{J}_0(\mathcal{A})<0$ the 
effective temperature is negative, since
$f_T(\omega/\mathcal{J}_0(\mathcal{A}))=f_{-T}(\omega/|\mathcal{J}_0(\mathcal{A})|)$.
%we can define an effective temperature $T_{\rm eff}$ as 
%$1/(e^{\omega/\mathcal{J}_0(\mathcal{A})T}+1) 
%\equiv 1/(e^{\omega/T_{\rm eff}}+1)$ 
%($T_{\rm eff} = \mathcal{J}_0(\mathcal{A}) T$),
%so that $T_{\rm eff}$ becomes {\it negative}.
%That is, negative temperature distribution is realized.
One intuitive way to interpret this is that, as  
the band is flipped when $J$ changes sign due to the 
ac field, each fermion follows the change 
with no shift in the momentum (Fig.~\ref{inversion} for $T<0$). 
This is the essential mechanism for the negative $T$ distribution that we propose here.
%This kind of distribution quite characteristic to nonequilibrium appears
%due to a mismatch of the sign of hopping amplitudes between $J$ and $J_{\rm eff}$,
%and we cannot exclude the negative sign of $T_{\rm eff}$ by particle-hole transformation.

\begin{figure}[t]
  \begin{center}
    \includegraphics[width=8.5cm]{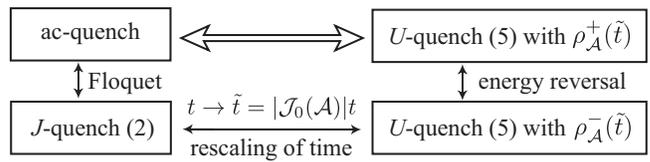}
    \caption{The ac/$U$-quench correspondence.}
    \label{logic}
  \end{center}
\end{figure}

Interacting systems driven by ac fields have also 
been theoretically studied for the Bose-Hubbard model \cite{EckardtWeissHolthaus2005,CreffieldMonteiro2006}, 
and the Falicov-Kimball model \cite{TsujiOkaAoki2009,TsujiOkaAoki2008}.
Let us assume that $J$ is suddenly quenched by the ac field as in Eq.~(\ref{bessel}). 
Then the density matrix time-evolves 
%in the presence of many-body interactions 
as $\rho(t)=e^{-it(-J_{\rm eff}\mathcal{H}_{K}+U\mathcal{H}_{I})}
\rho(0) e^{it(-J_{\rm eff}\mathcal{H}_{K}+U\mathcal{H}_{I})}
=e^{\mp i\tilde{t}(-J\mathcal{H}_{K}+U_{\rm eff}\mathcal{H}_{I})}
\rho(0) e^{\pm i\tilde{t}(-J\mathcal{H}_{K}+U_{\rm eff}\mathcal{H}_{I})} 
\equiv \rho_{\mathcal{A}}^\pm(\tilde{t})$,
where the upper (lower) sign corresponds to 
$\mathcal{J}_0(\mathcal{A}) >(<) 0$,
and time is rescaled as $\tilde{t}=|\mathcal{J}_0(\mathcal{A})|t$. 
After that, the $J$ quench is translated to the interaction quench \cite{EcksteinKollarWerner2009} as
\begin{align}
  U
    \to
      U_{\rm eff}=U/\mathcal{J}_0(\mathcal{A}).
  \label{Uquench}
\end{align}
For $\mathcal{J}_0(\mathcal{A})<0$, the corresponding $U$ quench
evolves in time with $\rho_{\mathcal{A}}^-(\tilde{t})$ where the phase 
rotates in the reverse direction. We can 
relate $\rho_{\mathcal{A}}^-(\tilde{t})$ with the normal time evolution $\rho_{\mathcal{A}}^+(\tilde{t})$ 
via an energy reversal; i.e., the signs of all the quantities (except for time) that have dimension of energy 
are reversed. For example, energy reversal maps 
%$D(t)$ to $D(t)$, 
%D appears here for the first time, so you have to define it ?
%Why do you have to evoke D here ?
kinetic energy $E_{\rm kin}(t)\equiv -i\sum_{\bm k}\epsilon_{{\bm k}-{\bm A}(t)}G_{{\bm k}\sigma}^<(t,t)$ 
to $-E_{\rm kin}(t)$. We 
summarize the relation between ac quench and $U$ quench in Fig.~\ref{logic}.

The ac/$U$-quench correspondence, which we numerically demonstrate in the following, 
has the intriguing consequence that for $\mathcal{J}_0(\mathcal{A})<0$ the ac field effectively switches the many-body interaction 
from repulsive to attractive ($U_{\rm eff}<0)$ (\ref{Uquench}).
In addition, it implies that the system thermalizes to
a negative-temperature canonical distribution:
according to the thermalization hypothesis \cite{RigolDunjkoOlshanii2008, EcksteinKollarWerner2009}, 
the (nonintegrable) Hubbard model evolving with normal $\rho_{\mathcal{A}}^+(\tilde{t})$
thermalizes after a $U$ quench in the long-time limit with a positive temperature $T_{\rm eff}(>0)$.
Since $\rho_{\mathcal{A}}^-(\tilde{t})$ is related to $\rho_{\mathcal{A}}^+(\tilde{t})$ via the energy reversal  (Fig.~\ref{logic}),
the ac/$U$-quench correspondence implies that 
the system driven by the ac field should finally thermalize 
with a negative temperature, $-T_{\rm eff}/|\mathcal{J}_0(\mathcal{A})|$.

{\it Numerical result.---}
%So far the argument strongly relies on the assumptions of ac/$U$-quench correspondence.
%that (i) the scaling (\ref{bessel}) holds for correlated fermions as well, 
%and (ii) the interacting system
%{\it thermalizes} to a negative temperature distribution (\ref{rho}).
To verify the interaction switching implied by the ac/$U$-quench correspondence in the dynamical 
band-flipping in interacting systems,
we have numerically computed the 
time evolution of the system using the nonequilibrium DMFT
\cite{FreericksTurkowskiZlatic2006}, which can treat the dynamics of infinite systems  with the local correlations from the many-body interactions fully taken into account.  
In the DMFT calculation, we restrict ourselves to paramagnetic phases without charge order.
%In the DMFT calculations, we search for solutions without symmetry breaking.
The hopping is scaled with dimension $d$ as
%the band dispersion and group velocity 
%read $\epsilon_{\bm k}=-2J\sum_{i=1}^d \cos k_i$ 
%and $v_{\bm k}=2J\sum_{i=1}^d \sin k_i$ 
$J=J^\ast/2\sqrt{d}$ ($d\to\infty$) \cite{GeorgesKotliarKrauthRozenberg1996a}.
In the following we take $J^\ast$ as the unit of energy and write $J^\ast$ as $J$. 
%A summation in terms of ${\bm k}$ is done through joint density of states
%$\rho(\epsilon,v)=e^{-\epsilon^2-v^2}/\pi$ \cite{TurkowskiFreericks2005}.
Even when fermions are initially weakly interacting, the effective interaction (\ref{Uquench}) plunges into 
the strong-coupling regime 
around the zeros of $\mathcal{J}_0(\mathcal{A})$, so that 
we need an impurity solver that is valid
for both weak and strong interactions. Here we employ the continuous-time QMC method
based on a weak-coupling expansion \cite{WernerOkaMillis2009},
which is numerically exact within statistical errors. 
%For solving Kadanoff-Baym equations in the self-consistent condition of DMFT, 
%we basically follow Eckstein {\it et al}. \cite{EcksteinKollarWerner2010}

A simple measure of whether the interaction is 
repulsive or attractive is the double occupancy, $D(t)=\langle n_{j\uparrow}(t)n_{j\downarrow}(t) \rangle$.  
Figure \ref{dbleocc} displays the time evolution of the double occupancy 
for various amplitudes $\mathcal{A}=K/\Omega$ with 
fixed $\Omega=2\pi$, $U=1$, and $T=0.1$.  
$D(t)$ starts from the equilibrium value, 
which is, for the repulsive interaction, smaller 
than the noninteracting value, $\langle n_{j\uparrow}\rangle \langle n_{j\downarrow} \rangle = 0.25$.  
For $\mathcal{A}<\mathcal{A}_1$ (left panel of Fig.~\ref{dbleocc}),
$D(t)$ decreases with $t$, 
%with the large-$t$ value more suppressed for larger $\mathcal{A}$, 
which is natural since the effective interaction (\ref{Uquench})
is enhanced. 
%\textbf{CHECK THIS}
For large effective interaction, the double occupancy shows $2\pi/U$-periodic collapse-and-revival oscillations \cite{EcksteinKollarWerner2009}, while 
%and particles try to avoid to come to the same site more than the initial condition. 
the fast oscillations with frequency $2\Omega$ %seen in $D(t)$
come from a nonlinear effect of the ac field.

The double occupancy starts to behave in a dramatically different manner 
as we plunge into the $\mathcal{J}_0(\mathcal{A})<0$ regime, 
i.e. $\mathcal{A}_1<\mathcal{A}<\mathcal{A}_2=5.520\dots$ (right panel of Fig.~\ref{dbleocc}): 
$D(t)$ steeply increases after  the ac field is switched on, 
and even goes ``beyond'' the noninteracting value 0.25.
This implies that fermions prefer a large double occupancy, 
evidence that the many-body interaction indeed turns into an attraction.  
When $\mathcal{J}_0(\mathcal{A})$ returns to positive 
for $\mathcal{A}>\mathcal{A}_2$, $D(t)$ again becomes 
smaller than 0.25,
and the effective interaction goes back to repulsive.  
%The sign of $\frac{1}{4}-D(t)$
%roughly follows that of $\mathcal{J}_0(\mathcal{A})$, so that $U_{\rm eff}(\mathcal{A})$ continues to oscillate
%between positive and negative.

\begin{figure}[t]
  \begin{center}
    \includegraphics[width=80mm]{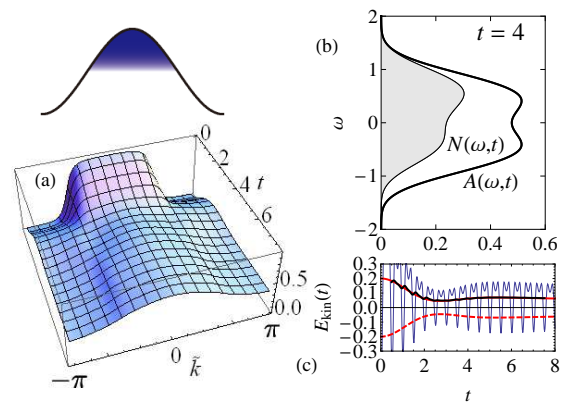}
    \caption{(a) The momentum distribution, 
    (b) the spectral function (thick curve), the occupation (shaded region), and 
    (c) the kinetic energy for an ac quench with $\mathcal{A}=4$, $\Omega=2\pi$ and $U=1$
    (oscillatory curve, along with a time-average in black), as compared with 
    the corresponding $U$ quench (dashed curve, along with a sign-inverted plot).
    The top left inset schematically shows a negative-$T$ situation.}
    \label{distribution}
  \end{center}
\end{figure}

To endorse the ac/$U$-quench correspondence (Fig.~\ref{logic}) quantitatively, 
%we pose a question: does the ac-quench resemble 
%an interaction quench \cite{EcksteinKollarWerner2009}, 
%where we suddenly change $U$ according to Eq.~(\ref{Uquench}) 
%instead of suddenly switching on the ac field.  
%Here we define a new density matrix conjugate to 
%$\rho(t)={\rm T} e^{-i\!\int_0^t d\bar{t}H(\bar{t})} \rho(0) e^{i\!\int_0^t d\bar{t}H(\bar{t})}$ as
%$\bar{\rho}(t)={\rm T} e^{i\!\int_0^t d\bar{t}H(\bar{t})} \rho(0) e^{-i\!\int_0^t d\bar{t}H(\bar{t})}$
%(${\rm T}$ is a time-ordering operator).
%Note the sign difference in the exponents between $\rho_{\mathcal{A}}^+$ and $\rho_{\mathcal{A}}^-$.
%Thus the ac-quench should correspond
%to $U$-quench (\ref{Uquench}) with the time evolution given by $\rho(t)$ ($\mathcal{J}_0(\mathcal{A})>0$)
%or $\bar{\rho}(t)$ ($\mathcal{J}_0(\mathcal{A})<0$). A physical quantity $\bar{O}(t)$ described by $\bar{\rho}(t)$
%that has a dimension of energy is equivalent to $O(t)$ described by $\rho(t)$ with the sign reversed
%in $t>0$: for example, $\bar{D}(t)=D(t)$, the total energy $\bar{E}(t)=-E(t)$, and
%$\bar{A}(\omega,t)=A(-\omega,t)$.
%\textbf{SHOULD REPLACE $K(t)\rightarrow E_{kin}(t)$}
we plot $D(t)$ [solid curves in Fig.~\ref{dbleocc}, plotted as a function of the rescaled time $t=\tilde{t}/|\mathcal{J}_0(\mathcal{A})|$] and 
$E_{\rm kin}(t)$ [dashed curve in Fig.~\ref{distribution}(c)]
%compare ac-quench (curves with error bars) and 
for the $U$ quench calculated with $\rho_{\mathcal{A}}^+(\tilde{t})$. 
%are superposed as solid curves.  
For $D(t)$, we can see that the $U$-quench results 
agree with those for the (nonoscillatory components of) the ac quenches to a surprisingly good accuracy. 
The accuracy is especially excellent for $\mathcal{A}\gtrsim 4$, while 
for $\mathcal{A}\lesssim 3$ the small differences seen 
between the $U$ and ac quench come from 
the fact that it takes 
%finite time for the ac field to renormalize $J$ in the first few cycles.  
a few cycles for the ac field to renormalize $J$.
%For  $\mathcal{A}\gtrsim 4$, $J$ is instantly rescaled even when $\mathcal{J}_0(\mathcal{A})<0$.
%Thus the correspondence becomes almost perfect except for the oscillating components.  
%For $K(t)$ with 
For $\mathcal{J}_0(\mathcal{A})<0$ we again confirm 
the ac/$U$-quench correspondence in the kinetic energy if 
we consider the energy reversal $E_{\rm kin}(t)\to -E_{\rm kin}(t)$ in the $U$ quench 
[a sign-inverted plot of the dashed curve in Fig.~\ref{distribution}(c)]
and time-average $E_{\rm kin}(t)$ over one cycle in the ac quench (black curve).
%(with appropriate energy reversal for $\mathcal{J}_0(\mathcal{A})<0$ because of the sign reversal
%in $\rho_{\mathcal{A}}^-$).
%In the language of Floquet theory, this fact suggests that 
%after sufficient time has passed 
%in the Floquet matrix $\hat{H}_{mn}$
%is block-diagonalized with no transition among different Floquet modes, 
%$(0,0)$-block obeys the scaling (\ref{bessel}).
%The correspondence not only gives a clear-cut indication of the attractively 
%converted interaction, but also accounts for the detailed behavior 
%of the temporal evolution of the system:  the double occupancy 
%in the ac-quench is seen to exhibit an overshooting and 
%restoration in Fig.~\ref{dbleocc}, which should correspond 
%to the $2\pi/U$-oscillation known to exist  in $U$-quench \cite{EcksteinKollarWerner2009}.

As for the population inversion, we can also obtain its direct evidence 
by calculating the (gauge-invariant) momentum distribution function \cite{TsujiOkaAoki2008} 
$f(\tilde{\bm k},t)\equiv -iG_{\tilde{\bm k}}^<(t,t)$
($\tilde{\bm k}(t)={\bm k}+{\bm A}(t)$) in Fig.~\ref{distribution}(a), 
and $A(\omega,t)$ and $N(\omega,t)$ \cite{Hanning} in Fig.~\ref{distribution}(b)
for the same $U$, $\Omega$, and $T$ as in Fig.~\ref{dbleocc}, with $\mathcal{A}=4$ 
[for which $\mathcal{J}_0(\mathcal{A})<0$]. 
We plot $f(\tilde{\bm k},t)$ along a typical slice \cite{Momentum} in the Brillouin zone.
%, which can be defined even for many-body 
%systems by the equations above eqn.(\ref{spectrum},\ref{occupation}).  
%$f(\tilde{k},t)$ is a top view of the distribution in a band picture (Fig.~\ref{inversion}), 
%whereas $N(\omega,t)$ is a side view (Fig*). 
In the initial state, the particles distribute around $-\frac{\pi}{2}<\tilde{k}<\frac{\pi}{2}$ with a blurred Fermi surface.  
As the ac field is turned on, the distribution is gradually smeared out, 
but the main population continues to 
stick to $-\frac{\pi}{2}<\tilde{k}<\frac{\pi}{2}$.
This sharply contrasts with what has been observed in BECs \cite{Lignier2007}: 
there, the system adiabatically follows the lowest energy states with no population inversion (Fig.~\ref{inversion} for $T>0$) 
with the peak of $f(\tilde{k},t)$ shifted by $\pi$ in the flipped band \cite{EckardtHolthaus2007}.
However, our results indicate that, with sufficiently fast ramp up of the ac field in fermion systems,
the system does not relax to lower energy states, but the population can be inverted. 
The population inversion is more directly seen in $N(\omega,t)$ [Fig.~\ref{distribution}(b)],
where fermions tend to occupy higher energy states ($\omega>0$),
%energy levels above $\omega=0$ tend to be occupied rather than those below $\omega=0$.
justifying our picture (Fig.~\ref{inversion} for $T<0$).
This kind of drastic change of the dynamics as a function of the parameter 
is reminiscent of phenomena observed in nonlinear mechanics.
%looks analogous to phenomena observed in nonlinear mechanics.
It may thus be interesting to try to connect bifurcation theory
\cite{GuckenheimerHolmes}
to the band-flipping phenomena found here.

\begin{figure}[t]
  \begin{center}
    \begin{tabular}{cc}
%      \hspace{-3mm}
      \includegraphics[width=40mm]{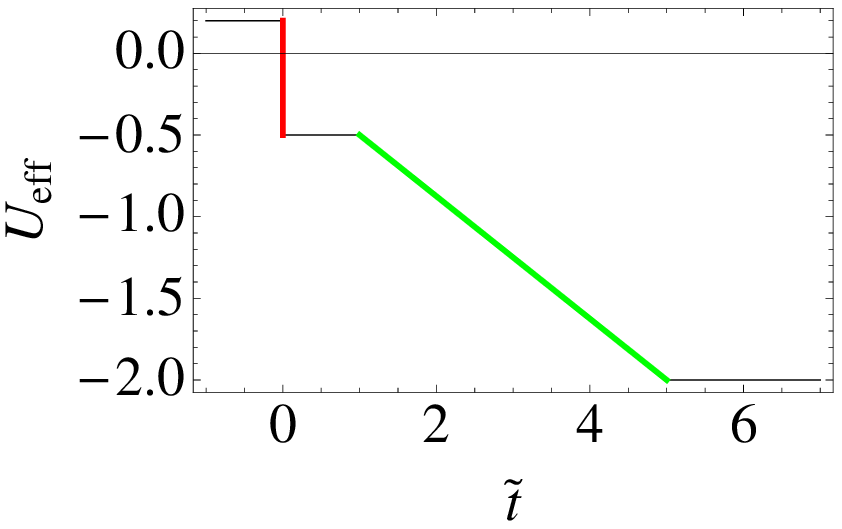}
%      \hspace{-6mm}
      \includegraphics[width=42mm]{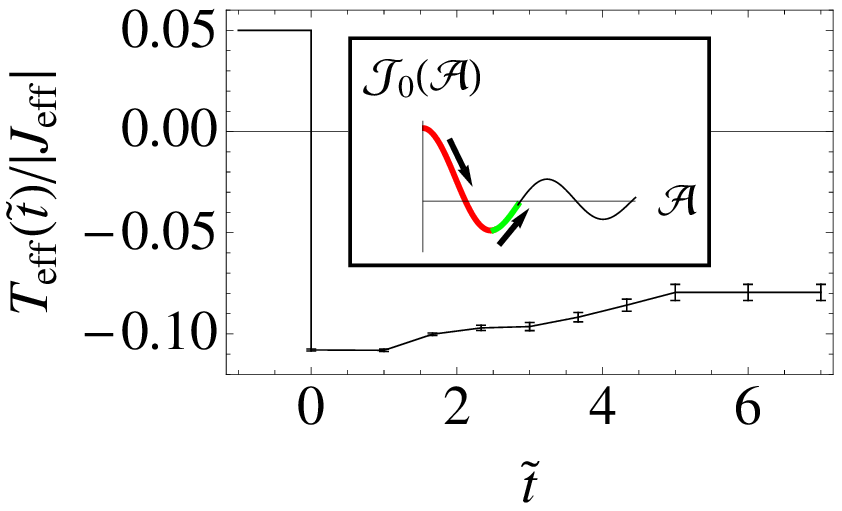} \\
    \end{tabular}
    \caption{An example of a multistep quench (left panel), 
    and the corresponding evolution of $T_{\rm eff}(\tilde{t})$ (right).
    The inset indicates the path for the ac quench on 
$\mathcal{J}_0(\mathcal{A})$.}
    \label{multi-step}
  \end{center}
\end{figure}
%{\it Possibility of dynamical superconductivity ---} 
{\it Suppression of heating ---} 
Now that the ac field is found to effectively convert a repulsive interaction into 
an attraction, one may contemplate the possibility of ac-induced 
superconductivity (SC), assuming that the system thermalizes. The obvious interest is that it is far easier 
to realize SC in attractive systems with $s$-wave pairing 
%with a BCS gap function 
%$\Delta=-U\sum_{\bm k}\Delta\tanh(\epsilon_{\bm k}/2T)/(2\epsilon_{\bm k})$ 
than in repulsive ones where the SC gap has to be anisotropic, 
so that the ac drive may provide an advantageous as well as 
novel avenue for SC.  
%For the ac-driven case, the mean-field gap equation reads
%$\Delta=-U\sum_{\bm k}\Delta\tanh(\mathcal{J}_0(\mathcal{A})\epsilon_{\bm k}/2T_{\rm eff}^\star)
%/(2\mathcal{J}_0(\mathcal{A})\epsilon_{\bm k})$, which can have a 
%solution for $\mathcal{J}_0(\mathcal{A})<0$ and $T_{\rm eff}^\star<0$
%even though $U>0$. 
%(or in other words, the total energy after $U$-quench should be less than equilibrium one for $T=T_c$).  
%This implies that one has to suppress the heating associated with the ac-quench 
%during application of ac fields to keep 
%below $|T_{\rm eff}^\star/J_{\rm eff}|\lesssim 0.1$.  
However, a simple, sudden quench 
%such as $U=1\to -2$ 
would heat the system to, e.g., 
$|T_{\rm eff}/J_{\rm eff}|=1.83\pm 0.02$ for $\mathcal{A}=4, \Omega=2\pi$, and 
$U=1$ (Fig.~\ref{distribution}),
which is much larger than the critical temperature for the attractive Hubbard model 
($T_c/J \sim 0.1$ for $U\sim -2$ \cite{MicnasRanningerRobaszkiewicz1990}).
So the crucial question is whether we can avoid such a heating.  

A smooth change in $U_{\rm eff}$ from positive to negative 
might seem desirable, 
but this is unfortunately impossible, since $|U_{\rm eff}/U|$ 
has to cross a singularity at $|\mathcal{J}_0(\mathcal{A})|=0$.  
Instead, we propose here a ``multistep quench'' 
(an example is shown in the left panel of Fig.~\ref{multi-step}). 
(i) We start from a weak interaction.
%(e.g., $U=0.2$), 
(ii) This is followed by a sudden quench (to avoid the above problem) to an attractive but still weak interaction. 
%$U_{\rm eff} (=0.2\to -0.5$).
The reason we start from a small $U$ is that 
a larger jump of $U_{\rm eff}$ in a sudden quench tends to cause a higher temperature.  
(iii) After the sudden quench we employ a smooth ramp to amplify $U_{\rm eff}$. 
%$(=-0.5\to -2.0)$. 
%we allow a thermalization to occur 
%The ac/$U$-quench correspondence and thermalization hypothesis provide a necessary condition for the 
%ac-driven SC:
%$|T_{\rm eff}/J_{\rm eff}|<T_c/J$ where $T_c$
%is the critical temperature for the attractive Hubbard model 
%($T_c \sim 0.1J$ for $U\sim -2$) \cite{MicnasRanningerRobaszkiewicz1990}.

In the right panel of Fig.~\ref{multi-step} we plot 
$T_{\rm eff}(\tilde{t})$ for one example of the multistep $U$ quench
($U_{\rm eff}=0.2 \to -0.5 \to -2$), which is estimated 
by equating the total energy of the system at each rescaled $\tilde{t}$ 
with the one for an equilibrium system with temperature $\tilde{T}$. 
%Only the final value of $\tilde{T}$ can be interpreted as $T_{\rm eff}/|J_{\rm eff}|=-\tilde{T}/J$.
For the example displayed here $|T_{\rm eff}(\tilde{t}=5)/J_{\rm eff}|=0.080\pm 0.004$ with an initial $T/J=0.05$, 
which accomplishes a temperature lower than $T_c/J\sim 0.1$.
%evaluate $|T_{\rm eff}^\star/J_{\rm eff}|=0.083\pm 0.003<T_c/J$, which is significantly below $T_c\simeq 0.1J$ required for SC.
%To summarize, an optimum route for the dynamical SC are: 
%(i) a small initial $U(\lesssim 0.1)$ followed by (ii) a multi-step (sudden + smooth) ac-quench with (iii) an initial $T<T_c$.  
%Since the anisotropic SC in the repulsive case 
%tends to have a critical temperature $\sim 0.01J$,  the attractively converted SC 
%with $T_c \sim 0.1J$ should provide an advantageous as well as 
%novel avenue for SC.  
%As for the particle concentration 
%Here we have only considered 
%the half-filled case, at which SC is degenerate with
%a charge density wave in the attractive Hubbard model. 
%If we go away from half-filling, to which our argument is straightforwardly applicable, SC should become dominant.
%Another point is that 
%one may think that for a symmetry-unbroken state
%to relax to a symmetry-broken state an excess energy must 
%be dissipated to the environment,
%which can destabilize the population inversion.  
%If one assumes the 
%thermalization hypothesis even for symmetry broken states,
%the excess energy is consumed for thermalizing in an isolated system
%keeping the symmetry broken, but this is an interesting future problem.

Finally we mention the stability and experimental feasibility of the population inversion. 
Once the negative-$T$ distribution is realized, it does not relax to a lower energy state
due to energy conservation as long as
%To maintain the stability of the population inversion it is imperative that 
the system is isolated from the environment with no energy dissipation.
If fermions are coupled to other degrees of freedom, the distribution 
should start to collapse to the
normal one with a positive $T$ with a decay rate determined by the strength of the coupling to the environment. 
One good candidate that can avoid this difficulty 
is a system of cold fermionic atoms trapped in an optical lattice, which is 
a virtually ideal, isolated single-band system \cite{Mosk2005}.  
There, a conversion from attractive to repulsive interactions, which is equally 
feasible, may also be interesting. 
For electron systems, possible candidates are superlattices or arrays of nanostructures 
designed to realize a single-band well separated from other bands. 
%In an ideally isolated system, our results imply that population inversion survives 
%even in the presence of interactions. In particular, with a continuous application of the ac field
%the total energy (not shown here) increases first and soon saturates with residual oscillations,
%meaning that at some point the ac field stops supplying energy to the system.
%Once population is inverted with the saturation, the stability of the distribution
%is guaranteed by the conservation of total energy.

%{\it Conclusion ---}
%We have shown that the sudden application of an ac field 
%to a lattice fermion system can
%dynamically flip its band structure, resulting in a negative temperature distribution, and that 
%the repulsive interfermion interaction can be effectively converted 
%into an attraction. Using the ramp-up protocol proposed here, 
%the effective temperature for the thermalized state can be suppressed below the $T_c$ of 
%the superconducting phase in the attractive Hubbard model.
%One can possibly use the effect to generate superconductivity.

%Acknowledgments.
We acknowledge illuminating discussions with M. Eckstein, T. Esslinger, and L. Tarruell.
N.T. was supported by the Japan Society for the Promotion of Science.

\bibliographystyle{apsrev}
\bibliography{floquet}

\begin{thebibliography}{21}
\bibitem[{Han()}]{Hanning}
\bibinfo{note}{To obtain $A(\omega,t)$ and $N(\omega,t)$ we Fourier-transform
  short time data using Hanning function.}
\bibitem[{Mom()}]{Momentum}
\bibinfo{note}{The slice is taken as $f(\tilde{k},t)=f(\tilde{\boldsymbol
  k},t)| _{\epsilon_{\tilde{\boldsymbol
  k}}=-\cos\tilde{k},v_{\tilde{\boldsymbol k}}=\sin\tilde{k}}$.}
\end{thebibliography}

\end{document}